\documentclass[sigconf]{acmart}

\setcopyright{rightsretained}
\copyrightyear{2020}
\acmYear{2020}
\acmDOI{10.1145/3394171.3414444}
\acmISBN{978-1-4503-7988-5/20/10}

\acmConference[MM '20]{Proceedings of the 28th ACM International Conference on Multimedia}{October 12--16, 2020}{Seattle, WA, USA}
\acmBooktitle{Proceedings of the 28th ACM International Conference on
Multimedia (MM '20), October 12--16, 2020, Seattle, WA, USA}

\begin{document}

\title{Visual-speech Synthesis of Exaggerated Corrective Feedback}


\makeatletter
\def\authornote#1{%
\g@addto@macro\@authornotes{%
      \stepcounter{footnote}\footnotetext{#1}}%
}
\makeatother

\author{
Yaohua Bu$^{*1}$,
Weijun Li$^{*2}$,
Tianyi Ma$^{3,4}$,
Shengqi Chen$^3$,
Jia Jia$^{\dagger 3,4, 5}$,
Kun Li$^6$,
Xiaobo Lu$^1$
}

\authornote{Equal contribution}
\authornote{Corresponding author, jjia@tsinghua.edu.cn} 

\affiliation{
$^1$ Academy of Arts \& Design, Tsinghua University \\
$^2$ School of Information Science and Technology, Northeast Normal University \\
$^3$ Department of Computer Science and Technology, Tsinghua University \\
$^4$ Key Laboratory of Pervasive Computing, Ministry of Education \\
$^5$ Beijing National Research Center for Information Science and Technology\quad
$^6$ SpeechX Ltd.
}

\renewcommand{\authors}{Yaohua Bu, Weijun Li, Tianyi Ma, Shengqi Chen, Jia Jia, Kun Li, and Xiaobo Lu}
\renewcommand{\shortauthors}{\authors}

\settopmatter{printacmref=false, printfolios=false}


\begin{abstract}
To provide more discriminative feedback for the second language (L2) learners to better identify their mispronunciation, we propose a method for exaggerated visual-speech feedback in computer-assisted pronunciation training (CAPT). 
The speech exaggeration is realized by an emphatic speech generation neural network based on Tacotron, while the visual exaggeration is accomplished by \textbf{ADC Viseme Blending}, namely increasing \textbf{A}mplitude of movement, extending the phone's \textbf{D}uration and enhancing the color \textbf{C}ontrast. 
User studies show that exaggerated feedback outperforms non-exaggerated version on helping learners with pronunciation identification and pronunciation improvement.
\end{abstract}



\maketitle

\section{Introduction}

Due to the influence of language transfer \cite{strange1995speech, meng2007contrastive, flege1995second, cortes2005negative, odlin1989language},  learners tend to replace an unfamiliar phone in the second language by a phone from their first language(L1). It will cause many inconspicuous mispronunciations \cite{lyster1998negotiation} that L2 learners hardly notice or correct them. There are many kinds of corrective feedbacks \cite{badin2010visual, wong2011allophonic, yuen2011enunciate, liu2012menunciate, agarwal2019review, derwing2005second} to increase the awareness between L1 and L2 in CAPT, such as speech, articulatory animations, etc. However, currently few feedback methods focus on the learners’ intuitive to realize their mistakes. Thus, offering an identifiable and perceptible feedback is necessary for the development of proper pronunciation.

When the learners are having difficulties to realize their mispronunciation, teachers widely use the exaggerated method in teaching, such as speaking loudly and slowly to show the movements of mouth clearly \cite{ricard1986beyond}. Inspired by this, we propose a method for visual-speech synthesis of exaggerated feedback to point out user’s mispronunciation by using emphatic speech and exaggerated articulatory animation. The speech and visual exaggeration are respectively realized by end-to-end emphatic speech synthesis and ADC Viseme Blending. Then they are combined to form the visual-speech synthesis\cite{wong2011allophonic}. By this method, we can provide more distinguishable feedback to learners and correct their mispronunciations.
\section{Implementation}

\subsection{Speech Exaggeration}
\label{sec:speech}
\begin{figure}[h]
  \centering
  \includegraphics[width=\linewidth]{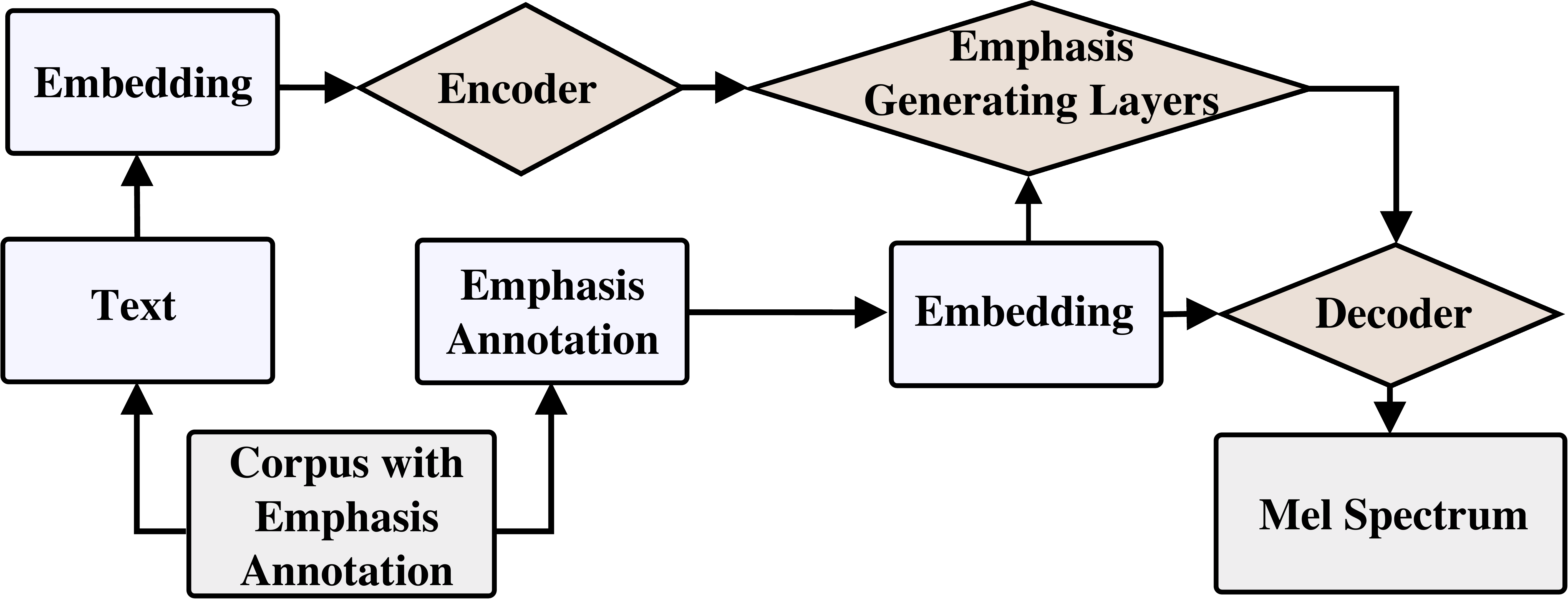}
  \caption{Architecture of speech exaggeration network}
  \label{fig:audio_network}
  \label{Architecture of speech exaggeration network}
  \Description{Architecture of speech exaggeration network}
\end{figure}

\begin{figure*}[t]
  \centering
  \includegraphics[width=1.0\linewidth]{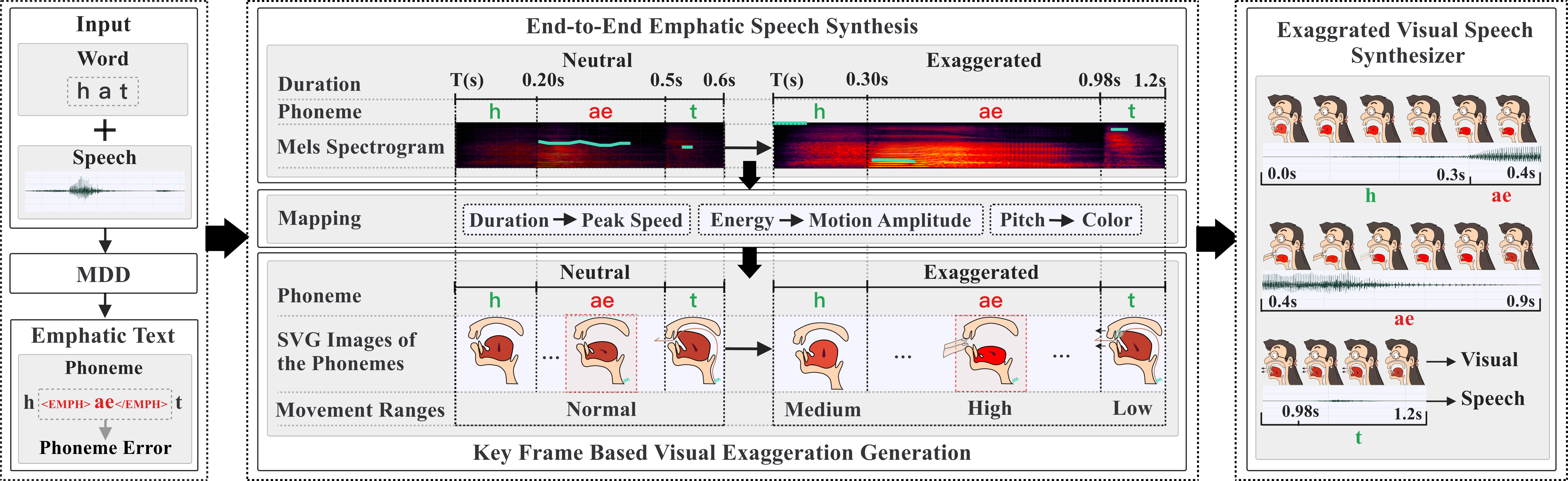}
  \caption{Flow chat of visual-speech synthesis of exaggerated corrective feedback}
  \label{fig:flow_chart}
  \Description{Flow chat of visual-speech synthesis of exaggerated corrective feedback}
\end{figure*}

To synthesize emphatic speech, 8000 text prompts with phone-level emphasis annotations are carefully designed, each containing one or more emphatic phones that the Chinese ESL speakers often mispronounce \cite{jia2014grading, ning2015hmm}. Two comparative speech utterances are recorded for each prompt: one with neutral intonation throughout the utterance and the other with exaggerated intonation with emphasis placed on the emphatic phones in the sentence.

We design a novel exaggerated speech generation neural network based on the previous Tacotron architecture \cite{shen2018natural, oord2016wavenet, oord2017parallel, wang2017tacotron}. The network synthesizes exaggerated speech from input pairs, each including a sentence text and an emphasis vector containing phone level exaggeration annotation of the corresponding sentence. Formally the network could be described as $W=G(T_{l\times h},A_{l\times 1})$, where $W$ is the sound wave with exaggeration, $T$ is the text input, $A$ is the annotation, $l$ is the length of a text,$h$ is the length of a text vector that represents a word and $G$ represents the whole network. 

The architecture of the network is shown in Figure \ref{fig:audio_network}. The pre-net embeds text with word vector containing phonetic level information, which are in the same shape and then processed separately in the encoder layer.
In the exaggeration generation layers, text and annotation vectors are combined, whose information are retained in output vector.
To enhance the effect of exaggeration annotation, it is mixed with the output of generation layer again in the decoder layer, whose output is transformed into Mel spectrum. The decoder also generates a sequence of the duration of each phoneme, which is later used by visual exaggeration.
Finally, WaveNet \cite{oord2016wavenet} is appended to convert the Mel spectrum to speech with exaggeration.







\subsection{Visual Exaggeration}
\label{sec:visual}

%

We generate the exaggerated version of speech animation mainly by \textbf{ADC} Viseme Blending, which means increasing the \textbf{A}mplitude of articulatory movement, the \textbf{D}uration of movement around the key actions\cite{zhao2013audiovisual} and color \textbf{C}ontract of movement. 

For each phone, we draw its corresponding visemes in SVG image according to the research on phonetics \cite{jones1922outline, carr2019english} and articulatory phonology \cite{browman1992articulatory, fougeron1997articulatory, zhang2017hearing} in four exaggeration levels, namely \emph{Normal}, \emph{Low}, \emph{Medium}, and \emph{High}. Each image includes tongue, chin, teeth, soft jaw and lips (see as Figure \ref{fig:flow_chart}). At each level, consonant and unit sounds each corresponds to one viseme, while each vowel has two. We then build a database for all these images.

To generate the animation of a specific phone sequence, we first sort out its corresponding viseme sequence, and obtain their duration from the network in Figure \ref{fig:audio_network}.
Then the acoustic features of phones are mapped to visual features, which is shown in Figure \ref{fig:flow_chart} and described below.

For different motion \textbf{A}mplitudes determined by the phonemes' energy, we choose SVG file of different exaggeration levels for each viseme, which are used as the key frames of the animation and then interpolated with non-linear functions for smoother and more natural transition between non-exaggerated and exaggerated phones. Also they are made more prominent by extending its \textbf{D}uration. Meanwhile, colors with higher \textbf{C}ontrast, purity and brightness are used to make the exaggerated key frames easier to identify. We also add auxiliary graphics, such as arrows and airflow, to help users to better understand the pronunciation through visualization.




\subsection{Visual-Speech Exaggeration Synthesis}

The flow chart describing the whole process of visual-speech synthesis of exaggerated corrective feedback could be seen in Figure \ref{fig:flow_chart}. We first decide the appropriate emphatic text by utilizing the Mispronunciation Detection \& Diagnosis (MDD) model\cite{li2016mispronunciation, li2015integrating, li2015rating}. The output text from MDD with emphatic marks is fed to the network introduced in Section \ref{sec:speech} to generate exaggerated speech feedback. Then we create the visual exaggeration in the format of SVG animation from the exaggerated speech by our proposed mapping rules as described in Section \ref{sec:visual}. Finally, speech and animation are combined to form the exaggerated visual-speech feedback.

\section{User Study}
\label{par:usrs}

Three experiments are conducted to compare the exaggerated (E) visual-speech feedback with the non-exaggerated (N) version.


First, we prepare 28 questions containing visual-speech animation of 14 pairs of easily mispronounced words (such as bed and bad), and randomly decide whether each animation should use E or N-feedback. 32 participants are asked to finish the questionnaire by distinguishing the two words in the pair. The average identification accuracy of E and N-feedback animation is 93.30\% and 75.45\% respectively, which proves the effectiveness of exaggerated feedback on pronunciation identification.

The second experiment is to verify whether exaggerated feedback could improve the learner's pronunciation accuracy. 14 participants with average English level are divided into N and E groups and take courses with N and E-feedback respectively. The result shows that the average increase of accuracy of E and N group is 19.92\% and 11.22\% separately, which indicates exaggerated feedback could better improve the accuracy of pronunciation.

In addition, in order to make exaggerated feedback more in line with learners' cognition and play a better teaching effect, we invite 29 learners and 20 professional teachers to rate the degree of exaggeration with average opinion score (MOS) method\cite{streijl2016mean}. For learners, the medium level of exaggeration gets the most three-point votes. For teachers, the medium level gets 86 three-point votes, which also prevails. Hence, the medium level is the most appropriate degree for exaggerated mispronounced phonemes.

\section*{Acknowledgement}
This work is supported by the state key program of the National Natural Science Foundation of China (No. 61831022) and the innovative research group project of the National Natural Science Foundation of China (No. 61521002).



\end{document}